\documentclass[12pt]{iopart}
\usepackage{iopams}

\usepackage{graphics}
\usepackage{graphicx}
\usepackage{epsfig}

\begin{document}

\newcommand\pb{\overline p}
\newcommand\qb{\overline q}
\newcommand\taubar{\overline\tau}

\title[Inhomogeneous Quasi-stationary States$\ldots$]{Inhomogeneous
Quasi-stationary States in
a Mean-field Model with Repulsive Cosine Interactions}

\author{F. Leyvraz$^\star$ \footnote[1]{Permanent address:
Centro de Ciencias Fisicas, UNAM, Apdo. Postal 48-3, 62251 Cuernavaca, Morelos,
Mexico}, M.-C. Firpo $^\star$\footnote[2]{Current address: 
Massachusetts Institute of Technology, Cambridge, Massachusetts 
02139-4307, USA}, S. Ruffo$^\star$\S}

\address{$^\star$ Dipartimento di Energetica ''Sergio Stecco'',\\
Universit\`{a} degli Studi di Firenze, Via Santa Marta, 3, I-50139 Firenze, Italy}
\address{\S\ INFM and INFN, Firenze}

\eads{\mailto{leyvraz@fis.unam.mx}, \mailto{firpo@mit.edu},
\mailto{ruffo@avanzi.de.unifi.it}}

\begin{abstract}
The system of $N$ particles moving on a circle and interacting via
a global repulsive cosine interaction is well
known to display spatially inhomogeneous structures of
extraordinary  stability starting from certain low energy initial
conditions. The object of this paper is to show in a detailed manner
how these structures arise and to explain their stability. By a
convenient  canonical transformation we rewrite the Hamiltonian in
such a way that fast and slow variables are singled out and
the canonical coordinates of a {\it collective mode} are
naturally introduced.
If, initially, enough energy is put in this mode,
its decay can be extremely slow. However, both analytical
arguments and numerical simulations suggest that these structures
eventually decay to the spatially uniform equilibrium state,
although this can happen on impressively long time scales.
Finally, we heuristically introduce a one-particle time
dependent Hamiltonian that well reproduces most of the
observed phenomenology.
\end{abstract}

\submitto{\JPA}
\maketitle

\section{Introduction}
\label{sec:intro}
Mean-field models, that is models in which all the particles interact
with equal intensity regardless of distance, have been the object of
considerable interest (see \cite{Review} for a review and the references therein).
One of the simplest models of this kind consists of $N$ particles moving on a circle
coupled globally by a cosine interaction\cite{Ruffowk,Antoni95}, with Hamiltonian:
\begin{equation}
H=\frac{1}{2}\sum_{i=1}^Np_i^2+\frac{J}{2N}
\sum_{i,j=1}^N\cos(\theta_i-\theta_j)~.
\label{eq:hamiltonian}
\end{equation}
It is sometimes called the Hamiltonian Mean-Field model (HMF). Here the variables
$p_i$ are the momenta conjugate to $\theta_i$, which
is the angle describing the state of the
$i$-th particle. As stated above, the strength of the interparticle
interaction does not depend on the distance
and all particles interact with all others.
Alternatively, one can think of this model as representing
a mean-field approximation to the classical XY model,
though this is certainly not a realistic model for a spin
system.
Its equilibrium statistical mechanics can
be treated exactly via the Hubbard-Stratonovich transformation,
as shown in~\cite{Antoni95,Latora99}. The attractive
case, corresponding to $J<0$, is shown to have a phase transition at a certain value
of the inverse temperature $\beta_c=2/J$, above which the stable state has
a {\it clustered} structure: that is, the particle density on the circle
shows a non-trivial profile.
Analogous results were also obtained using entropy maximization
methods~\cite{Ina93a,Ina93b} and, numerically, for a time discrete version
of model (\ref{eq:hamiltonian})~\cite{Kan92,Kan94}.
On the other hand, in the repulsive case ($J>0)$, no phase
transition is found~\cite{Antoni95}. This can be heuristically
justified through the following physical argument. First, note that Hamiltonian
(\ref{eq:hamiltonian}) can be rewritten as follows:
\begin{equation}
H=\frac{1}{2}\sum_{i=1}^Np_i^2+\frac{J}{2N}\left[\left(
\sum_{i=1}^N\cos\theta_i\right)^2+
\left(\sum_{i=1}^N\sin\theta_i\right)^2\right].
\label{eq:hamiltonian2}
\end{equation}
From this follows immediately that, at least if $N$ is an even
number, any state in which all spins appear in pairs
$(\theta_k, \pi+\theta_k)$ for all $k$ corresponds to
a ground state configuration of the system,
all of such states having the same energy.
Since the positions of half the angles can be chosen arbitrarily
without in any way affecting the energy, uniform ground states will be
entropically favoured, thus precluding the appearance
of a thermodynamically stable density modulation at low temperatures,
as is indeed rigorously established in Refs.~\cite{Antoni95,Latora99}.

It was therefore quite intriguing to find spatially inhomogeneous
solutions in the antiferromagnetic case~\cite{Antoni95,Kan94,Gang99,Dauxois00}.
Numerically, they have been found in molecular dynamics simulations
starting from certain initial conditions at low temperatures.
These so-called {\it bicluster} solutions
show two paired clusters in the density, that is, particles appear
preferentially in correlated positions $(\theta, \pi+\theta)$.
These biclusters were found to be extremely stable, though it is
of course never easy to decide on numerical ground
whether they are simply very long-lived or truly stationary.

This paper addresses the
following issues: in Section~\ref{sec:description}, we give a cursory
description of the numerical findings concerning the bicluster.
In Section~\ref{sec:anneau}, we show how the
$N$ particle system can be reduced to a system in which the particles
are only linked to each other through the common interaction
with a {\it collective mode}. We shall show that this phenomenon is reminiscent of
the wave-particle interaction mechanism that governs
a large amount of plasma physics phenomena~\cite{Antoni97,Timofeev97}. 
To make this picture
even clearer and more quantitative, we show in Section~\ref{sec:une-particule}
how this system can be further reduced to a one-particle time dependent
Hamiltonian. Finally, in Section~\ref{sec:analyse}, we perform a detailed analysis of
the one-particle model, showing that several features
of the bicluster can be explained in these terms. Finally, in
Section~\ref{sec:conclusions}, we present our conclusions.
\section{Description of the bicluster}
\label{sec:description}
In the following, we consider a system of $N$ particles interacting
according to Hamiltonian (\ref{eq:hamiltonian}) with a
positive value of $J$, corresponding to repulsion among
the particles. We shall consider almost
exclusively the following initial conditions: the initial
velocities are all identically zero, and the angles are
randomly distributed according to a uniform distribution on the
interval $[0,2\pi]$. Since this is not in general an equilibrium
position, the particles start moving and acquire
typical velocities of the order of $N^{-1/2}$.
We are therefore always dealing with the low-energy dynamics
of the system in the large $N$ limit. Under these circumstances, it is found numerically
that a bicluster forms, that is, the particle density
becomes inhomogeneous. The early stages of this formation were
studied analytically in Ref.~\cite{Barre}, using a convenient
zero-temperature limit of the Vlasov equation. These involve the rapid
formation of rather complex spatio-temporal inhomogeneities,
so-called {\it chevrons}, which, however, disappear on a fairly rapid
time scale. Since we are here primarily interested in
the long-time behaviour, these shall not concern us in
the following.

Once the initial stages of growth described in Ref.~\cite{Barre}
are over, an apparently stationary density $\rho(\theta)$ arises. The
time evolution of this density can be characterized by the following moments:
\begin{equation}
M_k(t)=\frac{1}{N}\sum_{m=1}^N \exp\left(ik\theta_m(t)\right).
\label{eq:def-moments}
\end{equation}
The second moment $|M_2(t)|$ is shown in Figure~\ref{fig006}
for a large system ($N=10^5$ particles). After some oscillations
corresponding to the chevron structures described in Ref.~\cite{Barre},
the second moment eventually reaches a constant value
of order one (though on the times shown in the figure, the increase is
not yet quite over).
Apart from quite rapid oscillations, which we always average over,
the density $\rho(\theta)$ settles down onto a provisionally stable profile, the
{\it bicluster}, which is shown in Figure~\ref{fig007}.

It is found on numerical evidence~\cite{Dauxois00} that
the time-averaged moments $M_k^{(0)}$ in this
{\it quasi-stationary state} are well approximated by the expression
\begin{eqnarray}
\left|M_0^{(0)}\right|&=&1\nonumber\\
\left|M_{2k}^{(0)}\right|&=&\frac{1}{|k|}\qquad(k\neq0),
\label{eq:moments-anal}
\end{eqnarray}
whereas they are zero for odd values of $k$.
This leads to the following analytical expression for the density
(again after averaging over the rapid oscillations):
\begin{equation}
\rho_0(\theta)=\frac{1}{2\pi}\left(1-\ln\left|2\sin\theta\right|~.
\right)
\label{eq:density}
\end{equation}
Here the moments $M_k^{(0)}$ and the density $\rho_0(\theta)$
are connected by the following relation:
\begin{equation}
M_k^{(0)}=\frac{1}{2\pi}\int_0^{2\pi}\rho_0(\theta)e^{ik\theta}
d\theta.
\end{equation}
The bicluster state does not last forever, at least for finite
values of $N$. Indeed, we have performed additional numerical
simulations of the
system and could observe the decay of the second moment
to its equilibrium value over a long time scale. Specifically,
for $N=50$, the relaxation of $|M_2(t)|$ is displayed in
Figure~\ref{fig008} for two different energy densities. We see that a significant
relaxation to equilibrium has taken place by a time of
approximately $4\cdot10^8$ for both energies.
These times rapidly increase as $N$ increases, so that,
in some sense, one might argue that the bicluster is some
kind of equilibrium state. This is related to the issue of
the non-commutation of the limits $t\to\infty$ and
$N\to\infty$ in mean-field or Vlasov-like systems
(see e.g. references \cite{LRR99,Firpo01} for a numerical
approach). On the other hand, the appearance
of a bicluster is closely linked to the peculiar choice of the initial
conditions: in particular, if instead of starting with zero velocities
and random positions, one starts with random velocities of the
same order as those that will eventually develop in the former
initial condition, no bicluster formation is observed. Together with
the fact that we are dealing with a very low-energy phenomenon in
a system with long-range forces, this leads us to suspect that
the bicluster is simply a phenomenon involving {\it metastable}
configurations.
Since it is well known that metastable states
become infinitely long-lived in the mean-field limit, this
point of view is in full agreement with the one involving
non-commuting long-time and thermodynamic limits.

\section{A simple equivalent model}
\label{sec:anneau}
\newcommand\barx{\overline{x}}
\newcommand\bary{\overline{y}}
\newcommand\half{\frac{1}{2}}
Our purpose in this section is to map Hamiltonian (\ref{eq:hamiltonian})
onto a system of the following form:
\begin{equation}
\tilde{H}=
\frac{\omega(\bphi)^2P^2}{2NJ}+\frac{NJX^2}{2}+
\half\sum_{k=1}^Np_k^2.
\label{eq:approx-hamiltonian}
\end{equation}
Here $P$ and $X$ stand for two canonically conjugate variables.
The vector $\bphi$ has coordinates $\phi_k$, conjugate
to the $p_k$ and $\omega(\bphi)$ is a smooth function which
will be computed explicitly in the following. Here the $\phi_k$
correspond to the angles of the original particles, whereas
$P$ and $X$ correspond to an explicit separation of a
{\em collective mode}. One therefore has the following
situation: the system consists of non-interacting particles
coupled to an oscillatory excitation, in a similar way as that
familiar in the wave particle models in plasma physics~\cite{Antoni97}.
Furthermore, as we shall see, the function $\omega(\bphi)$
is an average over all $\phi_k$, and therefore a slowly varying
function. Under these circumstances, one expects a very accurate
(though in general not exact) conservation of the adiabatic invariant
related to the energy stored in the collective mode.
As we shall see, if this invariant has an initial value
much larger than its thermal average value, a bicluster forms
and lasts as long as the invariant maintains its value.

To show this correspondence, we start by considering system
(\ref{eq:hamiltonian}) as a system of particles in two
dimensions, where all the particles are additionally constrained to
move on the unit circle. In order to deal with the constraint, it is
convenient to go to the Lagrangian formulation. One finds for the
Lagrangian corresponding to (\ref{eq:hamiltonian}):
\begin{eqnarray}
L&=&\half\sum_{k=1}^N(\dot{\barx}_{k}^2+\dot{\bary}_{k}^2)-\frac{J}
{2N}\left[\left(\sum_{k=1}^N\barx{}_k\right)^2
+\left(\sum_{k=1}^N\bary{}_k\right)^2\right]+\nonumber\\
&&+\sum_{k=1}^N\lambda_k(\barx{}_k^2+\bary{}_k^2-1).
\label{eq:cartesian}
\end{eqnarray}
This has the form of a non-linearly constrained {$N$-dimensional
harmonic oscillator.
We now separate the motion of the center of mass, since it
plays a peculiar role. We therefore introduce:
\begin{eqnarray}
X_1&=&\frac{1}{N}\sum_{k=1}^N\barx_k\qquad X_2=\frac{1}{N}
\sum_{k=1}^N\bary_k\nonumber\\
x_k&=&\barx_k-X_1\qquad y_k=\bary_k-X_2.
\label{eq:center-of-mass}
\end{eqnarray}
In these new coordinates the Lagrangian reads
\begin{eqnarray}
L&=&\half\sum_{k=1}^N(\dot x_k^2+\dot y_k^2)+\frac{N}{2}
\left[\dot X_1^2+\dot X_2^2-J(X_1^2+X_2^2)\right]+\nonumber\\
&&+\sum_{k=1}^N\lambda_k(x_k^2+y_k^2+2X_1x_k+2X_2y_k+X_1^2+X_2^2-1).
\label{eq:cartesian-com}
\end{eqnarray}
Going back to polar coordinates and making the following
approximations, which are valid in the limit
in which the formation of the bicluster is observed,
\begin{equation}
X_1,X_2\ll1\qquad\dot X_1,\dot X_2\ll1,
\label{eq:approx}
\end{equation}
one finally obtains for the Lagrangian, after some
manipulations shown in Appendix A,
\begin{equation}
L_0=\frac{\dot X^2}{2}\left(N+\sum_{k=1}^N\cos^2\phi_k\right)
+\frac{N}{2}X^2\dot\Phi^2+\sum_{k=1}^N\dot\frac{\phi_k^2}{2}-\frac{NJ}{2}X^2.
\label{eq:approx-lagrange}
\end{equation}
Here $X$ and $\Phi$ are related to $X_1$ and $X_2$ via
\begin{equation}
X_1=X\cos\Phi\qquad X_2=X\sin\Phi.
\label{eq:def-Phi}
\end{equation}
Let us now switch back to the Hamiltonian picture. The
Hamiltonian corresponding to (\ref{eq:approx-lagrange})
is exactly given by
\begin{equation}
H=
\frac{\omega(\bphi)^2P^2}{2NJ}+\frac{NJX^2}{2}+\frac{\Lambda^2}{2NX^2}
+\half\sum_{k=1}^Np_k^2.
\label{eq:approx-hamiltonian1}
\end{equation}
Here $P$ is conjugate to $X$ and $\Lambda$ to $\Phi$. As a final
remark, we note that $\Lambda$ is an exact constant of
the motion, namely the total angular momentum.
It can therefore always be set equal to zero
without loss of generality.  This therefore leads us to
(\ref{eq:approx-hamiltonian}).
The frequency function $\omega(\bphi)$ is given by
\begin{equation}
\omega(\bphi)=\sqrt{\frac{NJ}{N+\sum_{k=1}^N\cos^2\phi_k}
}.
\label{eq:def-omega}
\end{equation}
Hamiltonian (\ref{eq:approx-hamiltonian}) is very similar in form to those developed
in the theory of wave particle interactions in plasma physics.
However, it should be emphasized at this stage
that the system described by (\ref{eq:approx-hamiltonian})
is still an $N$-dimensional system. Contrarily to the reduction
proposed in \cite{Antoni97} where bulk and tail (i.e. suprathermal)
particles can be discriminated, our system is
`cold' so that all particles participate at the same level
to the collective mode.

A very appealing interpretation of (\ref{eq:approx-hamiltonian})
is the following. Consider a ring of mass one on which
$N$ beads of mass $1/N$ can move freely. If $X$ is taken
to be the amplitude of the torsional vibrations of such a ring
around the $y$ axis, and if no other motion
of the ring is allowed, then (\ref{eq:approx-hamiltonian}) describes
the motion both of the ring and the beads.
Consequently, if we start with particles at rest
and with an initial vibration entirely in the degree of
freedom associated to the ring, we will have the ring
oscillating first around its axis, and then driving the beads towards those
regions where the ring motion is least, that is, towards the
two points of the ring lying on the $y$ axis. Such
a structure corresponds exactly to what is observed for the bicluster.

We now proceed with the calculation. Our aim is twofold.
On the one hand, we show that equilibrium states do not contain
a density modulation. On the other, we display a mechanism capable
of leading to very long-lived inhomogeneous states.

Let us first consider the partition function of
the effective Hamiltonian (\ref{eq:approx-hamiltonian}).
The $p_k$ integrals factorize and one is left with
\begin{equation}
Z=\int_0^{2\pi}d\phi_1\cdot\ldots\cdot d\phi_N
\omega(\bphi)^{-1}
\int d{P}\,d{X}\exp\left(-\frac{\beta}{2}
(P^2+X^2)\right)
\label{eq:partition}
\end{equation}
This, however, is easily shown not to lead to any
density modulation in the angles, since
$\omega(\bphi)$ is of order one. We must therefore look
beyond statistical equilibrium, something already indicated
by the plausibility arguments mentioned in the Introduction
as well as by exact results.

To this end, we transform $X$ and $P$ locally to
action angle variables, that is, we transform to
\begin{eqnarray}
I&=&\half\left(
\frac{\omega(\bphi) P^2}{NJ}+\frac{NJ X^2}
{\omega(\bphi)}
\right)\nonumber\\
\psi&=&\arctan\frac{\omega(\bphi) P}{NJ X}~.
\label{eq:action-angle}
\end{eqnarray}
This transformation must be complemented by an approppriate
transformation of the $\phi_k$ to make it canonical. This is
computed and discussed further in Appendix B. The final hamiltonian
is
\begin{equation}
\tilde{H}=\omega(\bphi)I(\bphi)
+\half\sum_{k=1}^N\left(P_k+\frac{I(\bphi)\psi}{\omega(\bphi)}
\frac{\partial\omega(\bphi)}{\partial\phi_k}\right)^2,
\label{eq:approx-hamiltonian-3}
\end{equation}
where the $P_k$ are the new particle momenta after the canonical
transformation. They are given by (\ref{eq:oldp-newp}).
Note that the interaction
terms, which are linear in the $p_k$, all average to zero
in the fast variables $\psi$, and the others are of
order $1/N^2$. They can therefore be accurately taken into
account via a perturbative averaging approach.
The adiabatic theorem then states that the
action $I(\bphi)$ is conserved to
a high degree of precision and over long time scales.
We may therefore construct an approximately stationary
state as follows: set an arbitrary value for the action
$I$ and, using this now as an {\it external\/} parameter,
compute the resulting equilibrium state for the $\phi_k$.

It should be noted here that recently, using a completely different approach
involving multiple time expansion, close results have been derived by Barr\'{e}
\etal in Ref.~\cite{preprint}. We believe that our approach 
complements their work by seeking to provide some intuition for 
the physical mechanisms involved in these processes.

If we now take $I$ to have
given values in (\ref{eq:approx-hamiltonian-3}) and proceed to
compute the Gibbs partition function, we can still eliminate
the shifts in the $P_k$'s and then factor the resulting
integral out. We are then left with
\begin{equation}
Z=\int_0^{2\pi}d\phi_1\cdot\ldots\cdot d\phi_N
\exp\left[-\beta\omega(\bphi)I\right].
\label{eq:partition-approx}
\end{equation}
Since $\omega(\bphi)$ is of order one, it is necessary
for $I$ to be of order $N/\beta$ (that is,
$N$ times larger than predicted by equipartition) in order to
induce a density modulation of order one in the angles. On the other
hand, if such a modulation is induced, it is straightforward
to verify that it will indeed be a $\pi$-periodic modulation
directed along the direction in which the collective mode
oscillates, and therefore has most of the obvious properties observed
in the bicluster. Strictly speaking, however, the
microcanonical ensemble should be used in this calculation, which
may possibly open a broader range of possibilities, as
regions of negative temperature become accessible. For a
discussion of these issues, see \cite{preprint}.

As stated in Section~\ref{sec:description}, various initial conditions
may or may not lead to the formation of a bicluster. Among those that
are successful, there are
\begin{enumerate}

\item Initial conditions with equispaced angles and a sinusoidal
amplitude velocity perturbation: these initial conditions yield
a bicluster always, as they put in an extensive (though very small)
value for the action $I$ of the collective oscillation.

\item Initial conditions with random angles and zero initial velocities:
For these it is readily verified that the actions of the collective mode
are of order one, whereas the typical velocity of a particle decays as
$1/\sqrt N$. From this follows that the effective inverse temperature $\beta$
increases as $N$, thus yielding the appropriate result in
(\ref{eq:partition-approx}).

\end{enumerate}
\section{Reduction to a One Particle Model}
\label{sec:une-particule}
We here show how to further reduce the model to a one-particle
time dependent Hamiltonian. The assumptions involved are less well controlled
and the results we obtain only coincide with the original
model over fairly short times. However, in many features,
it will turn out that the description given
by this simplified model is extremely satisfactory.
To this end, we make the following remark: in the model
given, say, by (\ref{eq:approx-hamiltonian}), all particles
are coupled to one single collective mode. The dynamics of
this mode is determined both by an external harmonic force
and by the reaction forces to all the particles involved.
The system is therefore not a one-particle system. In order
to reduce it to such a system, we require an approximate
expression for the behaviour of the macroscopic mode.
Once this is given, a one-particle approximation for the
whole system immediately follows. From (\ref{eq:approx-hamiltonian}) 
one obtains
\begin{equation}
\ddot X=-\omega^2(\bphi)X~.
\label{eq:motion-collective}
\end{equation}
Let us now consider the picture of the bicluster developed
in Section~\ref{sec:anneau}. We had found there that the bicluster
was a quasi-stationary state in which the action of the collective
mode had a given (macroscopic) value and the $\bphi$ a
corresponding stationary distribution. From this picture and
(\ref{eq:motion-collective}) follows that, in the bicluster, the motion
of the collective mode can be well approximated by a simple harmonic
motion.

These observations are confirmed numerically in Figure 7 of
\cite{Dauxois00} which shows the phase-points of $X_{1}$ in
the dynamical regime where the bicluster forms. One can see that $X_{1}$
undergoes rapid oscillations of small amplitude around the origin, with its
phase only slowly changing in time. Note that, in reality, a slow
drift in the relative phase between $X_{1}$ and $X_{2}$ is actually observed. This is not,
however, due to a non-zero value of the
angular momentum, but rather to the fact that the description of
the motion as a two-dimensional harmonic oscillator is only approximate
and not valid over large times.

Armed with this knowledge concerning the dynamics of the collective
mode, we can proceed to analyze the behaviour of the particles.
We go back to the original $\theta_k$ variables. The equation
of motion following from (\ref{eq:hamiltonian2}) can now be rewritten
as follows:
\begin{equation}
\ddot\theta_k=J\left[X_{1}(t)\sin\theta_k-X_{2}(t)\cos
\theta_k\right].
\label{eq:one-particle}
\end{equation}
{From} this, using the approximate description of the collective mode
described above, we obtain the following one-particle approximation:
\newcommand\eps{\varepsilon}
\begin{equation}
\ddot\theta_k=\eps\cos\omega t\sin\theta_k.
\label{eq:effectivedyn}
\end{equation}
Here the small parameter $\eps$ is equal to $J|X_{1}|$. Note that,
by hypothesis, the amplitude of the collective mode $X_{1}(t)$ is always
small, so that this last approximation is in fact justified
(see (\ref{eq:approx})).
In the next section we analyze this model in greater detail numerically.
Before we do this, however, let us quickly show how this can be treated
perturbatively.
The dynamics (\ref{eq:effectivedyn}) is described by the following
Hamiltonian
\begin{equation}
H=\frac{p^2}{2}-\eps\cos\omega t\cos\theta.
\label{eq:one-particle-ham}
\end{equation}
Using standard averaging techniques, one finds in the limit
of small $\eps$ that the motion is well described by the
following effective Hamiltonian:
\begin{equation}
\overline H=\frac{P^2}{2}-\frac{\eps^2}{8\omega^2}\cos 2\Theta.
\label{eq:ham-eff-final}
\end{equation}
Here $\Theta$ is $\theta$, from which a small rapid oscillation
has been substracted and $P$ is the canonicallly conjugate momentum.
Explicitly
\begin{equation}
\Theta=\theta+\frac{\eps}{2}\left(
\frac{\sin(\omega t-\theta)}{(\omega-P)^2}-\frac{\sin(\omega t+\theta)
}{(\omega+P)^2}
\right)~.
\end{equation}
It follows therefore that the particle moves, at least in an average
sense, in an effective potential of the form $-\cos 2\Theta$, thereby
leading to an attraction to the two points zero and $\pi$.

\section{Analysis of the one-particle model}
\label{sec:analyse}
In this Section, we investigate numerically whether the dynamical features
of the original model (\ref{eq:hamiltonian}) with cold initial conditions
can be accounted for by the minimal effective model (\ref{eq:effectivedyn})
with one-and-a-half degrees of freedom derived in the preceding Section.
We first observe that we can set $\omega=1$ without loss of generality,
because this parameter can be absorbed in a trivial time rescaling.
We then introduce the following area-preserving map with time-step
$\tau$, that constitutes the lowest order symplectic algorithm that numerically
solves (\ref{eq:effectivedyn}),
\begin{eqnarray}
\theta_{n+1} &=&\theta_{n}+p_{n+1}\tau
\label{eq:mapping} \\ p_{n+1}
&=&p_{n}-\varepsilon \tau \cos \left( \left( n+1\right) \tau \right)
\sin \theta_{n}. \nonumber
\end{eqnarray}
In order to reproduce the conditions of the simulations performed
in~\cite{Dauxois00}, we consider a set of $N$ initial conditions featuring a
spatially homogeneous distribution of particles of zero momentum.
That is, we take for all $j$, $1\leq j\leq N$
\begin{equation}
\theta_{0}^{j}=j\frac{2\pi }{N} ~{\rm{and}} ~p_{0}^{j}=0.
\label{eq:inicond}
\end{equation}
We then compute the resulting $N$ trajectories, obtaining their superposed
phase-space plots of figures \ref{fig001} and \ref{fig003}.
Since $\omega=1$ the period $T$ of the rapid oscillations of the
collective variable $X_{1}$ takes the value $T=2\pi$. In
these simulations, we fix the time step $\tau =T/100=2\pi/100$
and $\varepsilon=0.05$. Figure~\ref{fig001}
shows the superposed phase-space plots of the $N$ trajectories
(\ref{eq:inicond}) during the initial stage from which
the bicluster structure emerges. All trajectories wind around
the elliptic fixed points $\theta=0,\pi$. In figure~\ref{fig002}
we plot some trajectories with initial value $|\theta_{0}|<\pi /2$, which
show rapid oscillations of period $T$ superposed on a much slower
oscillatory motion around the minimum in $\theta=0$ of the
time averaged potential $-\cos 2\Theta $ in Hamiltonian (\ref{eq:ham-eff-final}). The
large time plots of figures \ref{fig003} clearly reproduce the features of the
full HMF $N$-degrees of freedom phase-space plots (e.g. figure 4 in
reference \cite{Dauxois00}).
The bicluster structure exhibits an overall oscillation on the period $T$
around the zero momenta fixed points $\theta=0$ and $\theta=\pi $.

It is also possible, using the effective Hamiltonian
(\ref{eq:ham-eff-final}), to get at least a qualitative understanding
of the behaviour of the position and momentum density functions, in particular of
their singularities. Indeed, if we assume that all particles start from rest
and are randomly distributed on the unit circle, we find for
the density $\rho(\theta)$
\begin{equation}
\rho(\theta)= (2\pi Z)^{-1}\int_0^{2\pi}d\theta_0
\int_{-\infty}^\infty dp\delta\left[\overline H(\theta,p)
-\overline H(\theta_0,0)\right],
\label{eq:dens-theta}
\end{equation}
where $Z$ is the microcanonical phase space volume and where we
have neglected the difference between the transformed variables
and the original ones. The integral in (\ref{eq:dens-theta})
can be evaluated  explicitly to yield
\begin{equation}
\rho(\theta)={\cal N}K(\cos^2\theta),
\label{eq:dens-theta-1}
\end{equation}
where $K(m)$ is the complete elliptic integral of the first kind as a function
of the parameter $m$ and ${\cal N}$ is an appropriate normalization constant.
The properties of the approximate solution (\ref{eq:dens-theta-1}) are very
similar to the ones observed for the full $N$-body system, e.g. we can
reproduce the logarithmic singularities at $\theta=0,\pi$. The general shape is
also similar (see figure \ref{fig-momentdens}), but the agreement is not as
satisfactory as that of the empirical formula (\ref{eq:density}), shown in
figure \ref{fig007}.

A similar calculation for the momentum density yields
\begin{equation}
\rho(p)={\cal N}K\left(1-\frac{w^4}{16}\right)\qquad(|w|\leq2),
\label{eq:dens-p}
\end{equation}
where $w$ is the appropriately scaled momentum $p/(\eps^2/8\omega^2)^{1/2}$.
Again, this has roughly the right form (see figure \ref{fig-positidens}), but,
quantitatively, it is not quite satisfactory.

Model (\ref{eq:effectivedyn}) seems therefore to be a good representation of the
dynamics of the antiferromagnetic HMF at vanishing energy, even if
the assumption of a constant frequency $\omega:=2\pi/T$ is (slightly)
violated due to self-consistency in the real system.

Let us now show that, although simple, this model has a non trivial
dynamics. First note that KAM theorem, which guarantees the preservation
of quasiperiodic motions under small perturbations, cannot be applied
in its classical form to
this system. Renaming the time $t$ as a phase variable $\varphi $,
the dynamics
(\ref{eq:effectivedyn}) derives from the two-degrees of freedom
Hamiltonian
\begin{equation}
H\left(\theta,p,\varphi ,u\right) =\frac{1}{2}p^{2}+u-\varepsilon\cos
\omega\varphi\cos \theta
\label{eq:Hameffective}
\end{equation}
where the variable $u$ is conjugated to time $\varphi$ and
does not appear in (\ref{eq:effectivedyn}). KAM theorem requires a
nondegeneracy condition on the frequency vector $\bomega:=\left(
\partial _{p}H,\partial _{u}H\right) $ to ensure that a large set of actions
$\left( p,u\right) $ have ``sufficiently irrational''
frequencies. This condition is $\det
\left( \partial _{(p,u)}\bomega\right) \neq 0$. This is
trivially not satisfied here as $\bomega=\left(p,1\right)$.
Yet, within the theory of averaging,
KAM theory may be applied to
an averaged Hamiltonian exponentially close to (\ref{eq:ham-eff-final})
with the result that, for $\varepsilon$ small enough,
the Poincar\'{e} section of $\overline H$ is
filled up to a residue of exponentially small measure by invariant
curves that are close to the level
lines of $\overline H$~\cite{Neihstadt,Treschev}.
This appears to be confirmed numerically, although in a rather
unexpected fashion. Figure~\ref{fig004}
displays the enlargement of a long-time Poincar\'{e} plot with initial
conditions equally spaced on the $\theta$-axis and $\varepsilon =0.05$.
Over the time scales probed
by the simulation ($4\cdot10^7$ timesteps), the particle does
not visit the whole phase space,
but remains within a well-defined region in agreement
with KAM predictions. In this region, however, it
appears to lie on an extremely convoluted but smooth invariant
surface. Indeed, the Poincar\'e surface shows a dense
pattern of parallel lines on which the points lie. These
presumably represent the intersection of a very high order
torus with the Poincar\'e surface. Such behaviour is certainly
peculiar, but may be well related to the degenerate nature of the
unperturbed system. From this follows, in this particular case, that
no short periodic orbits can exist in this part of the phase space.

If this picture is correct, the system is not ergodic.
This then strongly suggests that, in this reduced one-particle
model, the bicluster structure lasts forever. The relevance of this
conclusion for the $N$ particle system, however, remains an open
question. It is in principle conceivable that, among the class
of initial conditions we consider, a non-vanishing measure
lies on some invariant surface of the $N$ particle system.
However, this certainly does not follow from our previous arguments
concerning adiabatic invariance, which only holds over very long,
but not over infinite times. Furthermore, the simulations discussed
in Section \ref{sec:description} speak against such a possibility.

\section{Conclusions}
\label{sec:conclusions}
Summarizing, we have found a physically appealing approximate description
of the low-temperature dynamics of the repulsive
Hamiltonian  mean-field model
in terms of free particles on a ring performing torsional
vibrations. The physical analog of the ring oscillation is
the collective plasmon oscillation of the variables
$X_1$ and $X_2$. The interaction between particles and ring
arises solely from the influence of the particle positions on the
moment of inertia of the ring. This provides a straightforward
interpretation of the clustering phenomena observed in
previous work: if the initial condition is such that the initial
ring oscillation strongly dominates the motion of the individual
particles, then a parametric instability sets in, driving
the particles towards the part of the ring which is at rest. This
creates, as shown in Section~\ref{sec:une-particule}, an effective potential
of the form $-\cos 2\Theta$ in which the particles move. This same picture
can also be used to obtain a description of the statistical mechanical
equilibrium. In this case, it is clear that no clusters form, either
in the canonical or in the microcanonical ensemble. Furthermore,
this effective potential allows to form a qualitatively correct
picture of both the particle density in position and in momentum
space, though clear discrepancies remain, showing that the
time-dependent nature of the problem is essential. Finally,
we studied the chaotic properties of the
time-dependent effective model. For a small, yet finite, modulus of
the collective plasmon variable, this model was shown to present
localization on very high order tori, and hence lack of ergodicity.
This localization is presumably inherent to the small dimension
of the system and might not occur in the finite $N$ particle
model. Indeed, even if the tori present in the effective
one-particle system actually survived in the $N$ particle
system, Arnold diffusion might still act as a process
that would eventually drive the system towards thermal equilibrium.

\ack
MCF thanks European
Commission for support through a Marie Curie individual fellowship Contract
No HPMFCT-2000-00596. This work was partially funded by grant IN112200
of DGAPA as well as CONACYT grant number 32173-E. Finally, we
acknowledge financial support from INFN, the University of Florence
and the MURST-COFIN00 project {\it Chaos and localization in quantum
and classical mechanics}.

\appendix
\section{Derivation of the Approximate Lagrangian}
We start from (\ref{eq:cartesian-com}). We now introduce polar
coordinates as follows:
\begin{equation}
x_k=\rho_k\cos\phi_k\qquad y_k=\rho_k\sin\phi_k.
\label{eq:define-polar}
\end{equation}
Here the angles $\phi_k$ are measured from the vector $(X_1,X_2)$.
From this follows readily via the cosine theorem, that
\begin{equation}
\rho_k^2+X^2-2X\rho_k\cos\phi_k=1,
\label{eq:rel-rho-phi}
\end{equation}
where $X$ is defined as the norm and $\Phi$ as the angle of the
vector $(X_1,X_2)$. Within the approximations
stated in (\ref{eq:approx}), this yields
\begin{equation}
\rho_k=1+X\cos\phi_k.
\label{eq:rho-phi}
\end{equation}
From this follows for the velocities
\begin{equation}
\dot\rho_k=\dot X\cos\phi_k-X\dot\phi_k\sin\phi_k=\dot X\cos\phi_k,
\label{eq:vel-rho}
\end{equation}
where again the last equality uses (\ref{eq:approx})
and $\dot{X_1},\dot{X_2}\sim\dot{\phi}_k$. The Lagrangian
(\ref{eq:cartesian-com}) now reads
\begin{equation}
L=\half\sum_{k=1}^N(\dot\rho_k^2+\rho_k^2\dot\phi_k^2)+\frac{N}{2}
\left[\dot X^2+X^2\dot\Phi^2-JX^2\right]
\label{eq:polar-com}
\end{equation}
Substituting (\ref{eq:rho-phi}) and  (\ref{eq:vel-rho}) into
(\ref{eq:polar-com}) and using systematically the approximations
(\ref{eq:approx}) yields the desired result.
\section{Transformation to Action-angle Variables}
In this appendix, we compute the canonical transformation
that generates  the local transformation to action-angle
variables (\ref{eq:action-angle}). We shall do this by computing
its generating function as follows: we first determine a function
$S_0(I,X;\phi_k)$ such that the conditions
\begin{equation}
\frac{\partial S_0}{\partial I}=\psi\qquad
\frac{\partial S_0}{\partial X}=P
\label{eq:first-S}
\end{equation}
are equivalent to (\ref{eq:action-angle}). This is simply the
generating function of the transformation to action-angle
variables of the harmonic oscillator, where the $\phi_k$ enter
merely as parameters. If we now define
\begin{equation}
S(I,X;\phi_k,P_k)=\sum_{k=1}^N P_k\phi_k+S_0(I,X;\phi_k),
\end{equation}
this determines in the usual way a canonical transformation
on the whole space. To compute it explicitly, note that $S_0$
depends only on the combination $\omega(\bphi)I$, so that
\begin{eqnarray}
p_k&=&P_k+\frac{I\psi}{\omega(\bphi)}\frac{
\partial\omega(\bphi)}{\partial\phi_k}\nonumber\\
&=&P_k+\frac{I\psi}{\omega(\bphi)}\cos\phi_k\sin\phi_k\sqrt{
\frac{NJ}{\left(1+\sum_{l=1}^N\cos^2\phi_l\right)^3}
}
\label{eq:oldp-newp}
\end{eqnarray}
Note that these correction terms are of order $1/N$.



\newpage

\begin{figure}
\centerline{\psfig{figure=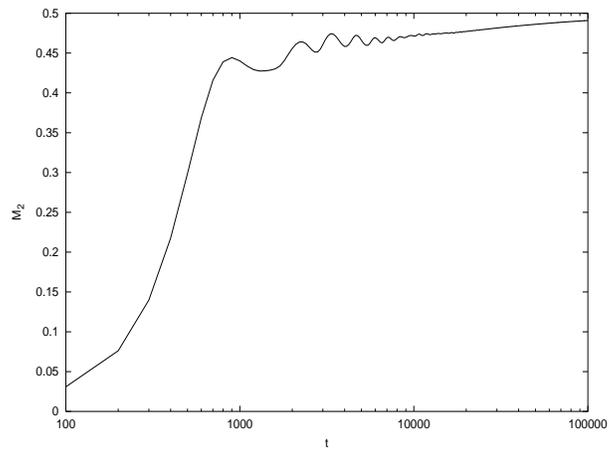,width=6cm,height=8cm,angle=-90}}
\vskip5mm
\caption{Time-dependence of $|M_2(t)|$ for a system of $N=10^5$
particles at short times.}
\label{fig006}
\end{figure}

\begin{figure}
\centerline{\psfig{figure=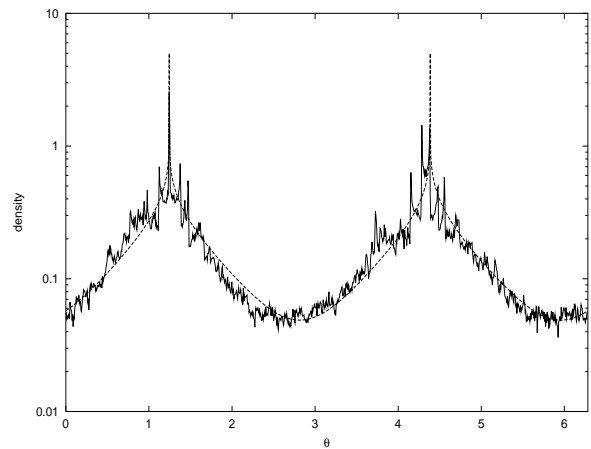,width=6cm,height=8cm,angle=-90}}
\vskip5mm
\caption{Bicluster density $\rho(\theta)$ for a system of $N=10^5$
particles. The full line is the result of numerical experiments,
the dashed line is the theoretical expression (\ref{eq:density}).}
\label{fig007}
\end{figure}

\begin{figure}
\centerline{\epsfig{figure=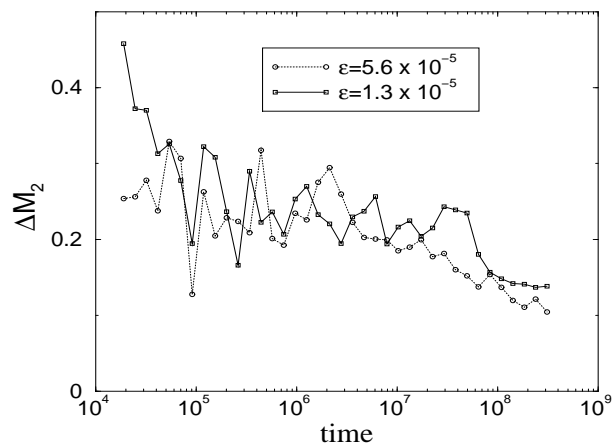,width=6cm,height=8cm,angle=-90}}
\vskip5mm
\caption{Decay of $\Delta M_2(t)$ ($|M_2(t)|$ minus the finite $N$ equilibrium
value) for two systems of 50 particles with different energy densities
$\varepsilon=H/N$ in log-linear scale.
To smooth out fluctuations, $\Delta M_2(t)$ has been averaged over
time intervals of exponentially growing size.}
\label{fig008}
\end{figure}

\begin{figure}
\centerline{\epsfig{figure=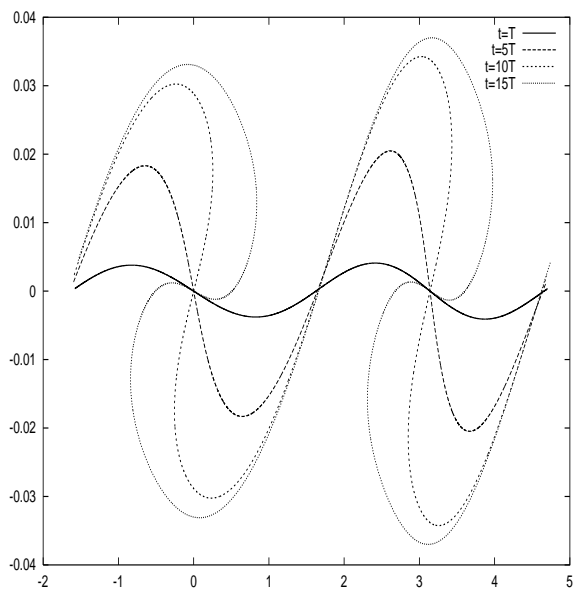,width=8cm,height=8cm,angle=-90}}
\vskip5mm
\caption{Superposition of one-particle phase-space plots
corresponding to the initial points (\ref{eq:inicond}),
during the initial stage of the bicluster formation.
The time unit is the short period $T=2\pi$.}
\label{fig001}
\end{figure}

\begin{figure}
\centerline{\psfig{figure=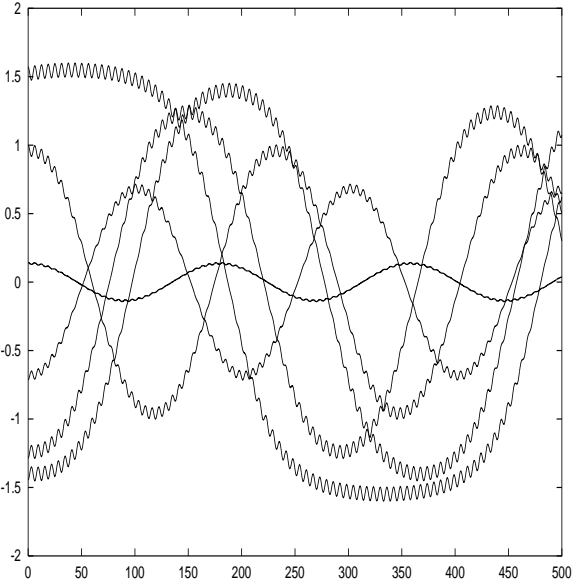,width=8cm,height=8cm,angle=-90}} 
\vskip5mm
\caption{Some trajectories $\theta(t)$ of the one-particle system.}
\label{fig002}
\end{figure}

\begin{figure}
\centerline{\hbox{\psfig{figure=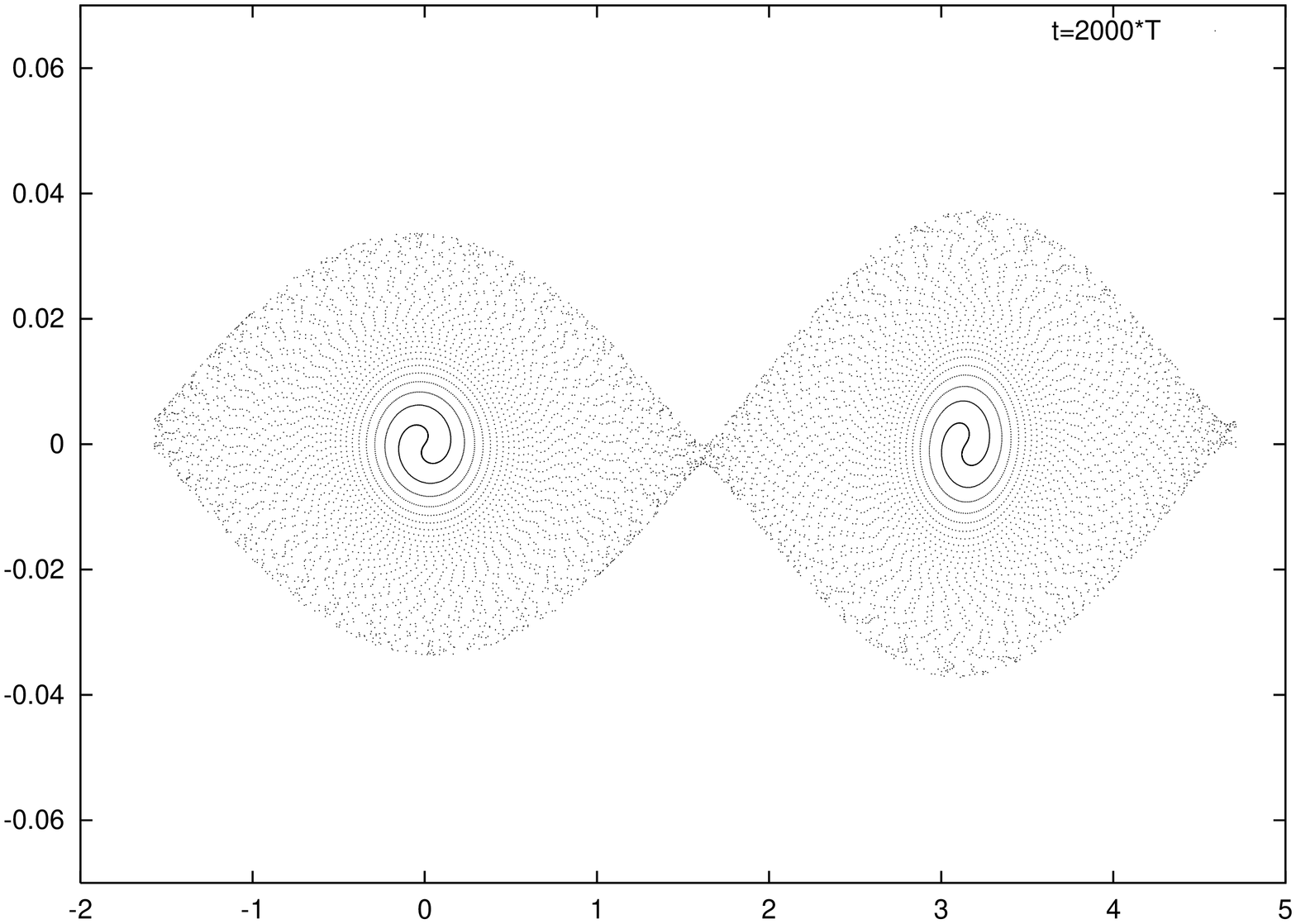,height=2in,width=2in,angle=-90}
\psfig{figure=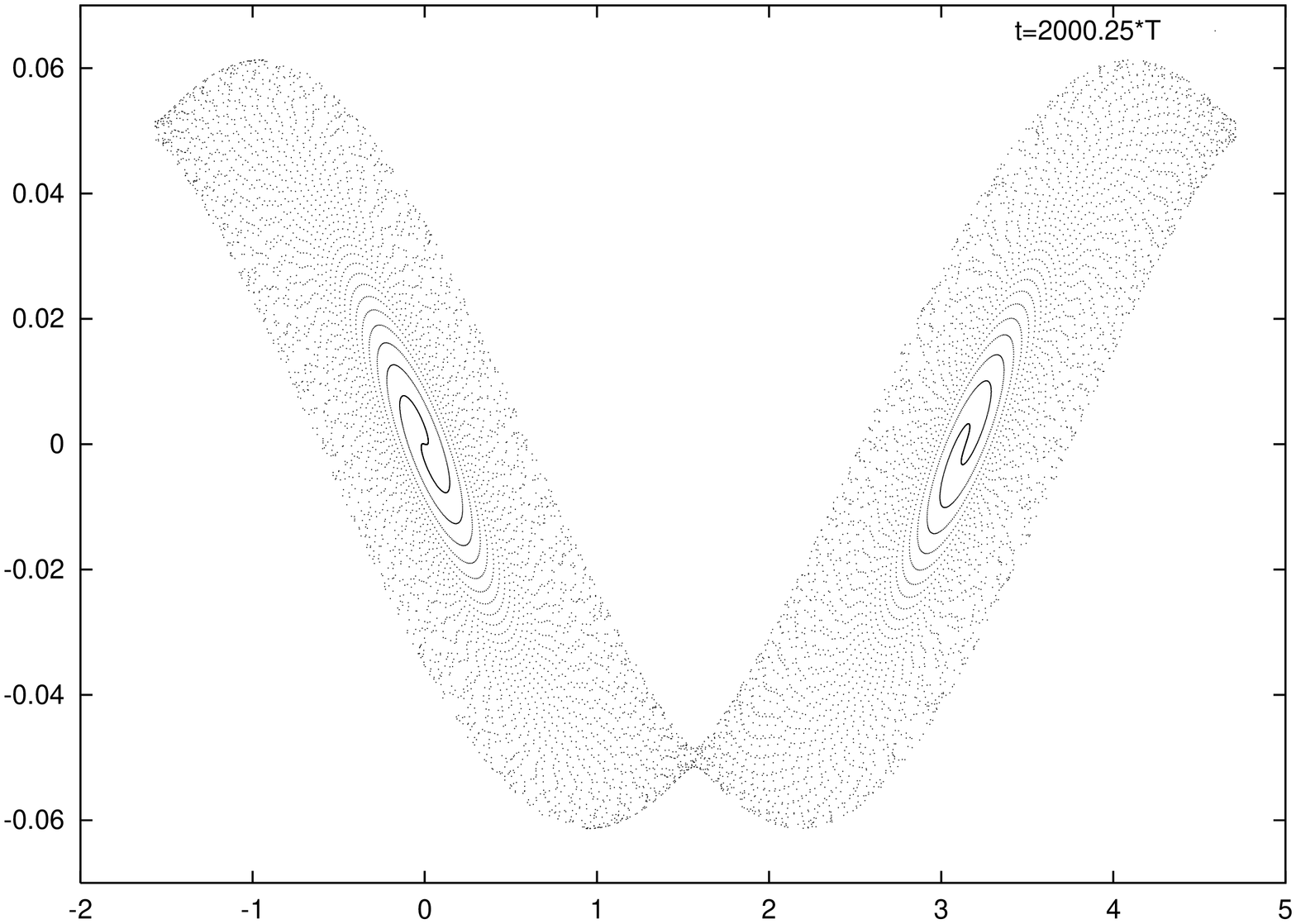,height=2in,width=2in,angle=-90}
}}
\centerline{\hbox{
\psfig{figure=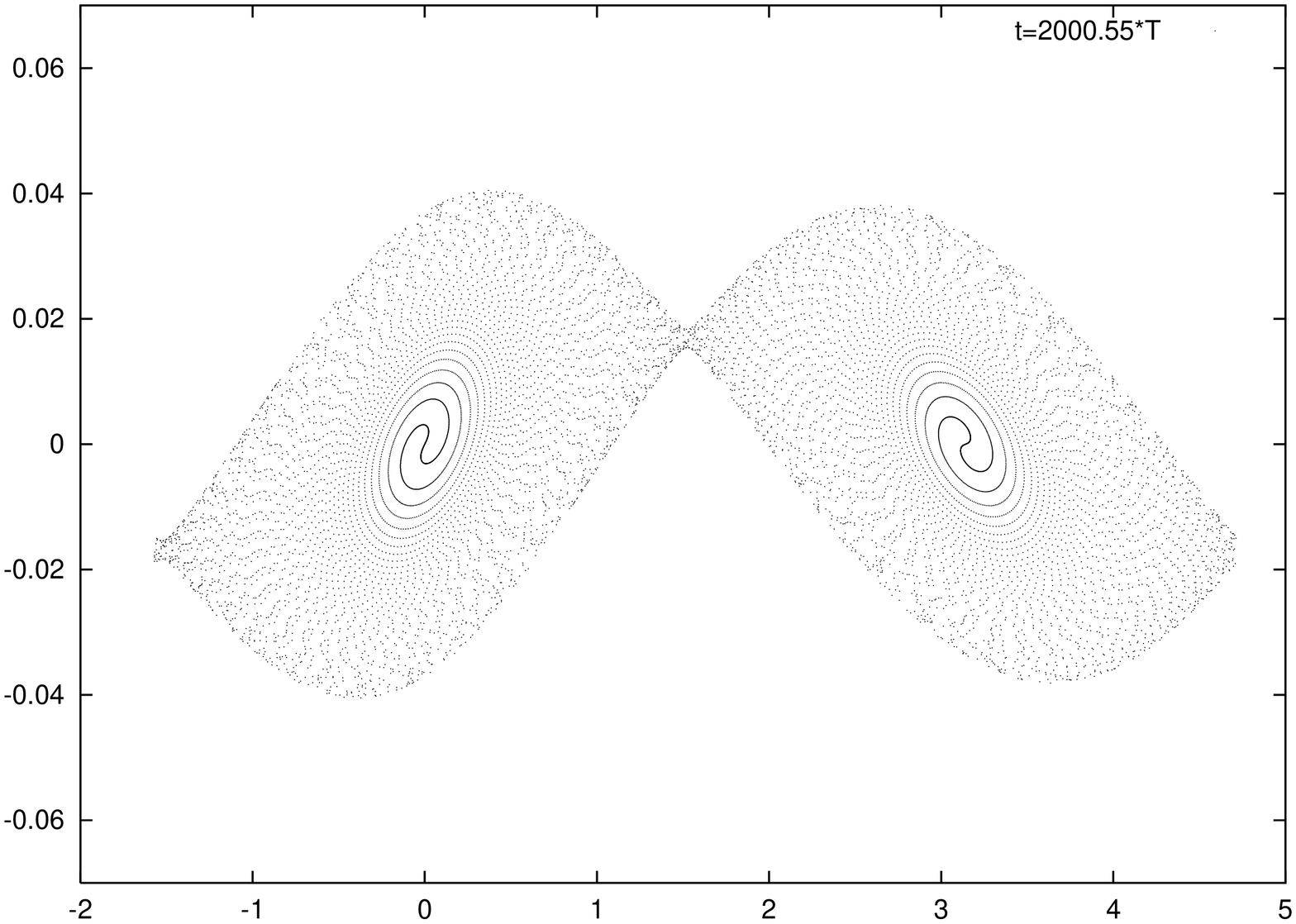,height=2in,width=2in,angle=-90}
\psfig{figure=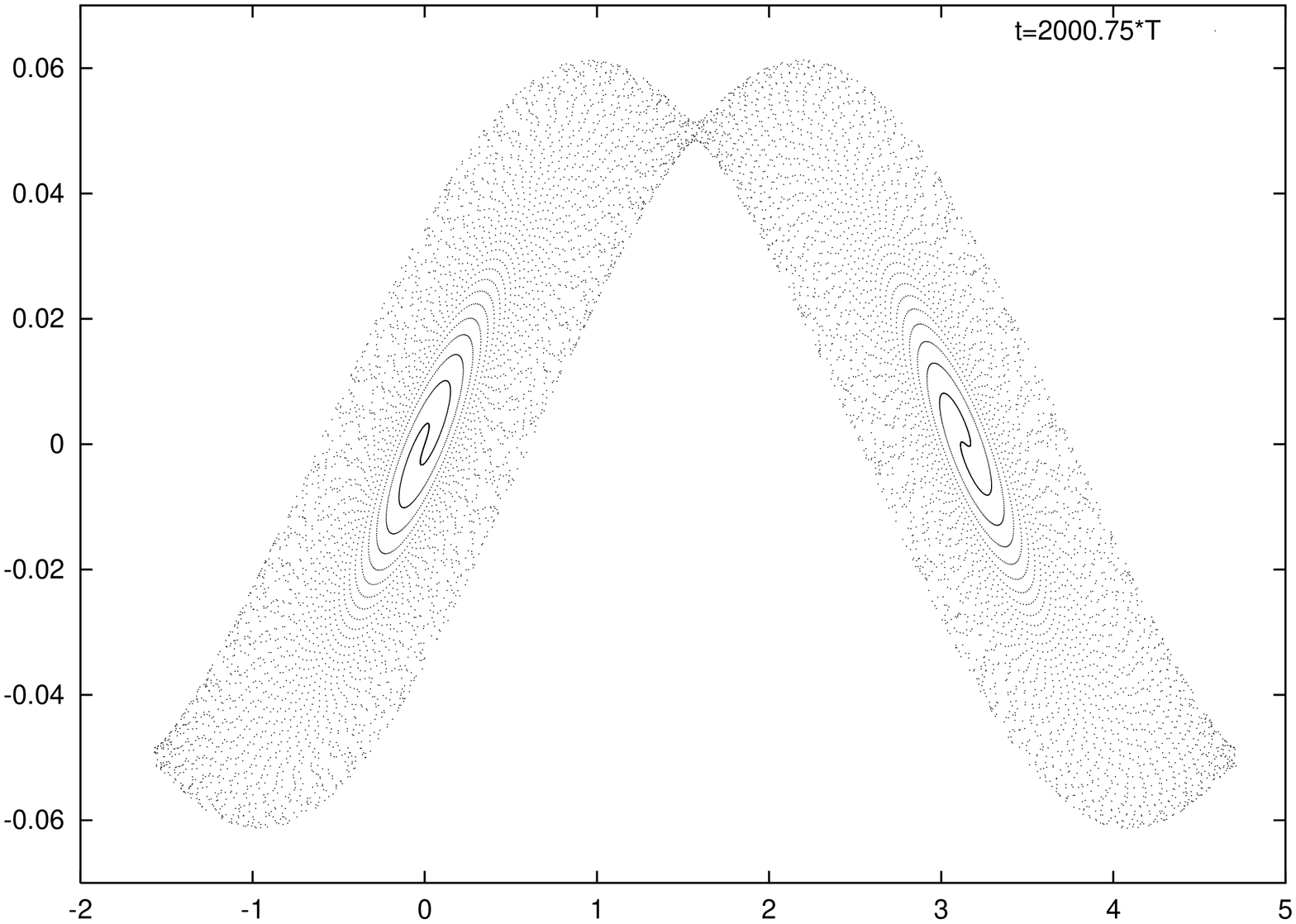,height=2in,width=2in,angle=-90}
}}
\vskip5mm
\caption{Four snapshots at long-time within the short period
$T=2\pi$ of the phase-space plots of the $N$ trajectories
(\ref{eq:inicond}). The particle positions have been
brought by periodicity to the interval
$[-\pi/2;3\pi/2]$.}
\label{fig003} 
\end{figure}

\begin{figure}
\centerline{
\psfig{figure=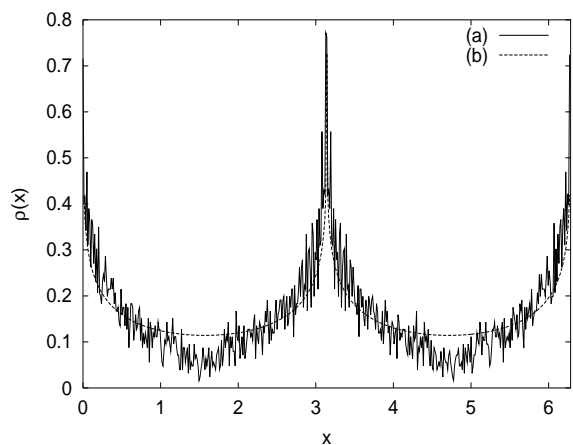,width=6cm,height=8cm,angle=-90}}
\vskip5mm
\caption{
Position density $\rho(\theta)$ at time $t=2000T$ of the bicluster
(a) from the numerical integration integration of (\ref{eq:mapping})
with the 'cold' initial conditions (\ref{eq:inicond}) in plain line and (b)
from the theoretical prediction (\ref{eq:dens-theta-1})
in dashed line.}
\label{fig-momentdens}
\end{figure}

\begin{figure}
\centerline{
\psfig{figure=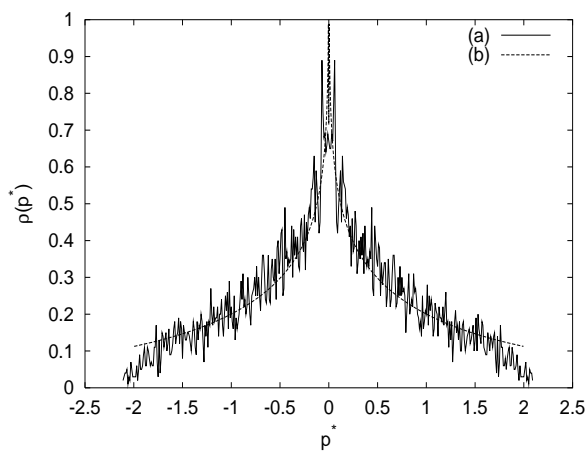,width=6cm,height=8cm,angle=-90}}
\vskip5mm
\caption{
Density of the rescaled momentum
$p^{*}=p/\sqrt{\varepsilon^2/(8\omega^2)}$ at time $t=2000T$ in the bicluster
(a) from the numerical integration integration of (\ref{eq:mapping})
with the 'cold' initial conditions (\ref{eq:inicond}) in plain line and (b)
from the theoretical prediction (\ref{eq:dens-p})
in dashed line.}
\label{fig-positidens}
\end{figure}

\begin{figure}
\centerline{
\psfig{figure=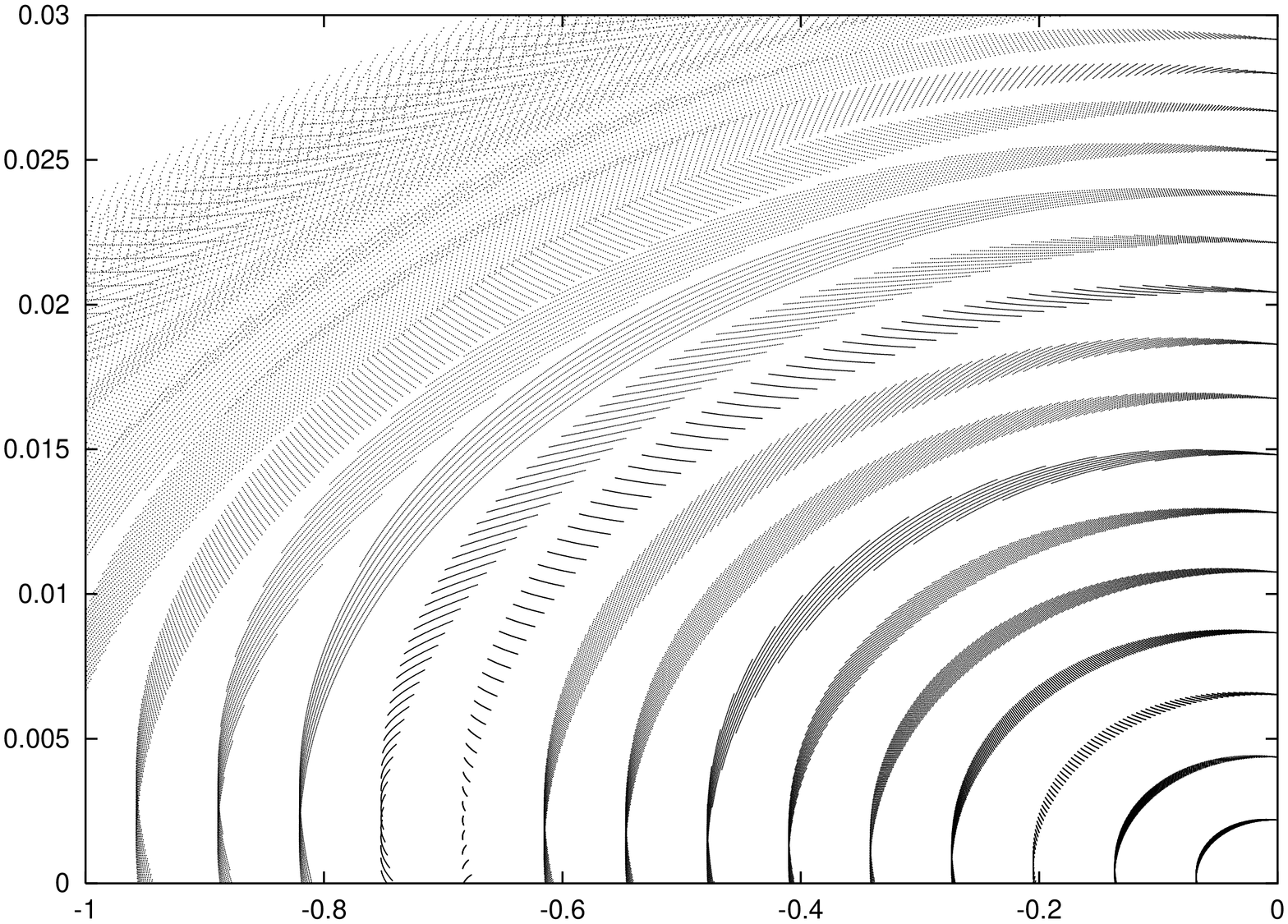,width=8cm,height=8cm,angle=-90}}
\vskip5mm
\caption{Magnification of a long-time Poincar\'{e} plots of system
(\ref{eq:effectivedyn}) corresponding to orbits of increasing
energy around the elliptic point $\theta=p=0$.}
\label{fig004}
\end{figure}

\end{document}